\documentclass[12pt]{article}
\usepackage{amsmath,amssymb}
\usepackage{latexsym,graphicx,verbatim}

\usepackage[usenames]{color}

\def\6#1{{\underline{#1}}}
\def\m6#1{{\underline{#1}\,}}

\newdimen\Tdim
\def\ispan{{\setbox0=\hbox{i}%
\Tdim\ht0\advance\Tdim\dp0\rule[-\dp0]{0pt}{\Tdim}}}
\def\jspan{{\setbox0=\hbox{j}%
\Tdim\ht0\advance\Tdim\dp0\rule[-\dp0]{0pt}{\Tdim}}}
\def\Tspan#1{{\setbox0=\hbox{#1}%
\Tdim\ht0\advance\Tdim\dp0\advance\Tdim.55ex\rule[-\dp0]{0pt}{\Tdim}\box0}}

\def\be{\begin{eqnarray}}
\def\ben{\begin{eqnarray*}}
\def\ee{\end{eqnarray}}
\def\een{\end{eqnarray*}}

\def\p{\partial}

\def\D{\mathcal{D}}

\def\=:{=\hspace{-.7em}\raisebox{1.1ex}{.}\hspace{.1em}\raisebox{-0.2ex}{.} }

\newcommand{\NF}{N_{\rm F}}

\newcommand {\beq}{\begin{eqnarray}}
\newcommand {\eeq}{\end{eqnarray}}

\makeatletter
\newcommand{\thetablename}{Table}
\def\fnum@table{\thetablename\ \thetable}
\makeatother

\begin{document}
\thispagestyle{empty}
\begin{flushright}
RIKEN-TH/2010-183\\
{\tt arXiv:1001.4320} \\
January, 2010 \\
\end{flushright}
\vspace{3mm}
\begin{center}
{\LARGE  
Trions in $1+1$ dimensions
} \\ 
\vspace{20mm}

{\Large
Minoru Eto
}
\footnotetext{
e-mail~address: \tt
meto(at)riken.jp
}
\vskip 1.0em
{\footnotesize
{\it Theoretical Physics Laboratory, Nishina Center, RIKEN, Saitama 351-0198, Japan
}
}
 \vspace{12mm}

\abstract{
We consider an Abelian BF-Higgs theory with $\NF=2$ Higgs fields in $1+1$ dimensions.
We derive a new BPS-like bound and find topological solitons with tri-charges (topological charge,
$Q$-charge and electric charge).
We call them ``trions."
}

\end{center}

\vfill
\newpage
\setcounter{page}{1}
\setcounter{footnote}{0}
\renewcommand{\thefootnote}{\arabic{footnote}}


\noindent{\bf Introduction}

Topological solitons are stable solutions  
of classical equations of motion in gauge theories and non-linear sigma models. 
The instantons in $4+1$ dimensions and the monopoles  
in $3+1$ dimensions are both well-known important solitons in the non-Abelian gauge theories. 
There are also solitons with di-charges. Electrically charged instantons are
called dyonic instantons \cite{Lambert:1999ua} and also monopoles with electric 
charge $Q_e$ are called dyons.
Interestingly, there are certain relations between solitons in higher dimensions
and lower dimensions. For example, it is known that the instantons/monopoles have 
many properties in common with the lumps/kinks in $2+1$/$1+1$ dimensions.
A counterpart of dyonic instanton/dyon is called $Q$-lump/$Q$-kink 
\cite{Leese:1991hr,Abraham:1992vb,Abraham:1992qv} which has
a conserved Noether charge $Q$, so-called $Q$-charge, instead of electric charges. 
Mass formulas of the dyonic instantons \cite{Lambert:1999ua} and $Q$-lumps \cite{Leese:1991hr} 
are indeed quite similar as
\beq
M_{\rm DI} = |Q_{\rm I}| + |Q_e|,\quad
M_{\rm QL} = |Q_{\rm L}| + |Q|,
\eeq
and that for the dyons and $Q$-kinks \cite{Abraham:1992vb,Abraham:1992qv} are given by
\beq
M_{\rm Dy} = \sqrt{Q_m^2 + Q_e^2},\quad
M_{\rm QK} = \sqrt{Q_{\rm K}^2 + Q^2}.
\eeq
Here $Q_{\rm I}, Q_{m}, Q_{\rm L}$ and $Q_{\rm K}$ stand for topological charges.
More direct relations can be seen when the monopoles are put into the Higgs phase.
In the Higgs phase monopoles are confined and attached to non-Abelian vortices.
In an effective world-sheet theory in $1+1$ dimensions, the monopoles are indeed identified with
kinks \cite{Tong:2003pz,Shifman:2004dr}. Similarly, instantons in the Higgs phase are 
identified with lumps \cite{Eto:2004rz} inside the vortex.
This way, in field theories, usually topological solitons have only two charges.

Linear BPS solitons with multiple charges were found in $2+1$ dimensions \cite{Kim:2006ee}.
Especially, the field theoretic supertube appears when its tension is cancelled
by a linear momentum density \cite{Kim:2006ee}.
Recently, a topological soliton with tri-charges was found
in $2+1$ dimensional Yang-Mills-Chern-Simons-Higgs theory.
It is called the dyonic non-Abelian vortex \cite{Collie:2008za}.
The topological vortices in Chern-Simons theories have electric charges
\cite{deVega:1986eu,Hong:1990yh,Jackiw:1990aw,Lee:1990eq}.
The dyonic non-Abelian vortex \cite{Collie:2008za} has not only
the topological and electric charges but also $Q$-charges. 

Inspired by the field theoretical supertube and dyonic non-Abelian vortex, 
we explore topological point-like solitons with tri-charges in
$1+1$ dimensions in this paper. We will consider $1+1$ dimensional BF-Higgs theory which can be obtained
by dimensional reduction from the Chern-Simons-Higgs theory in $2+1$ dimensions.
The topological/non-topological kink solutions were found in a similar BF-Higgs 
theory with a Higgs field \cite{Kao:1996tv}. In this paper we consider $\NF = 2$ Higgs fields. 
We derive a new BPS-like bound and find topological solitons which have tri-charges, 
topological, electric and $Q$-charges. 
We will call them ``trions" in $1+1$ dimensions.\\

\noindent{\bf Trions in BF-Higgs Theory}

We start with the so-called BF theory coupled with a real scalar field $N$ and
$\NF=2$ Higgs fields $\phi_1,\phi_2$
in $(1+1)$ dimensions with metric $\eta_{\mu\nu} = {\rm diag}(+,-)$
\beq
{\cal L}_{\rm BF} &=&
- \kappa N F_{01} 
+ \D_\mu \phi_1(\D^\mu \phi_1)^* + \D_\mu \phi_2(\D^\mu \phi_2)^*  - V_{\rm BF},\\
V_{\rm BF} &=& \sum_{i=1,2} \left[ \left(N - n_i\right)^2 |\phi_i|^2 
+ \left(\frac{1}{\kappa}\right)^2 
\left(|\phi_1|^2 + |\phi_2|^2 - v_i^2\right)^2 |\phi_i|^2
\right],
\label{eq:bf}
\eeq
where $\kappa$ is the BF coupling constant and $n_i$ is the Higgs mass.
We also introduced scalar coupling constants $v_i^2$.
The covariant derivative is $\D_\mu \phi_i= \p_\mu \phi_i + i A_\mu \phi_i$ and the field
strength is $F_{01} = \p_0 A_1 - \p_1 A_0$. Through out this paper, we assume $\kappa > 0$
and $v_i^2 > 0$.
We also consider non-degenerate parameters, $v_1 \neq v_2$ and/or
$n_1 \neq n_2$.
The mass dimensions of the fields and parameters are
$[\phi_i]=[v_i]=M^0$, $[A_\mu]=[N]=M^1$ and $[\kappa]=M^{-1}$.
Note that the BF-Higgs theory can be obtained by a suitable dimensional reduction
from the Abelian Chern-Simons theory in $2+1$ dimensions \cite{Kao:1996tv}.

For later convenience, 
let us define a gauge invariant object
\beq
M \equiv \frac{1}{\kappa}\left(|\phi_1|^2 + |\phi_2|^2\right).
\label{eq:def_M}
\eeq
Then the scalar potential can be rewritten
in a simple form
\beq
V_{\rm BF} = \sum_{i=1,2}
\left[ \left(N - n_i\right)^2 |\phi_i|^2 + 
\left(M - m_i\right)^2 |\phi_i|^2
\right],
\label{eq:bf_pot}
\eeq
with $m_i \equiv v_i^2/\kappa $.
Classical vacua correspond to $V_{\rm BF} = 0$. It follows that 
\beq
\text{Higgs}\ \left<1\right>&:&\ 
(N,M,\phi_1,\phi_2) = (n_1,m_1,v_1,0),
\label{eq:bf_higgs}\\
\text{Higgs}\ \left<2\right>&:&\ 
(N,M,\phi_1,\phi_2) = (n_2,m_2,0,v_2),
\\
\text{Coulomb}\ \left<0\right>&:&\ 
(N,M,\phi_1,\phi_2) = (\mathbb{R},0,0,0).
\label{eq:bf_coulomb}
\eeq
The $U(1)$ gauge symmetry is broken in the Higgs vacua and is unbroken in the Coulomb
vacua. The $N$ and $M$ are sufficient to specify the above vacua, so we consider the $NM$ plane.
The Higgs vacua exist only on the upper half $NM$ plane and the Coulomb vacua are
any points on the horizontal axes $(M=0)$.

The scalar field $N$ is a Lagrange multiplier. Its field equation gives a constraint
\beq
F_{01} = -\frac{2}{\kappa} \sum_{i=1,2}(N-n_i)|\phi_i|^2
\eeq
which determines distribution of the electric field.

The Bogomolnyi completion of the energy density is performed in a simple fashion
by making use of $M$ as
\beq
{\cal H} &=& 
\sum_{i=1,2}\left|\D_1 \phi_i + \left[(M-m_i)\cos\alpha - (N-n_i)\sin\alpha\right]\phi_i\right|^2 \nonumber\\
&+& \sum_{i=1,2}\left|\D_0 \phi_i - i \left[(M - m_i) \sin\alpha + (N-n_i)\cos\alpha\right]\phi_i\right|^2 \nonumber\\
&+& \p_1\left[
\frac{\kappa}{2}(N^2 - M^2) + \sum_{i=1,2} n_i J_{0i} + \sum_{i=1,2} m_i |\phi_i|^2
\right]\cos\alpha\nonumber\\
&+& \p_1\left[
\kappa MN + \sum_{i=1,2} m_i J_{0i} - \sum_{i=1,2} n_i |\phi_i|^2
\right] \sin\alpha,
\label{eq:bb_bf}
\eeq
where $\alpha$ is an arbitrary angle and we have used the Gauss' law
\beq
\kappa \p_1 N + \sum_{i=1,2} J_{0i} = 0,\quad
J_{0i} \equiv - i \left(\phi_i\D_0\phi_i^* - \D_0\phi_i\phi_i^*\right).
\label{eq:bf_gauss}
\eeq
The similar Bogomolnyi completion for the $\NF=1$ case with $m_1 =v^2$ and $n_1 = 0$ 
was found in \cite{Kao:1996tv,Kim:2006ee}.
Now we introduce topological charges
\beq
T_X \equiv \frac{\kappa}{2} \int dx\ \p_1(X^2),\qquad
\left(X=M,N\right).
\eeq
We also introduce the electric charge and the Noether charge so-called $Q$-charge
\beq
Q_e \equiv \int dx\, (J_{01}+J_{02}),\quad
Q \equiv \int dx\, \frac{J_{01} - J_{02}}{2}.
\eeq
The electric charge is expressed by the Gauss' law as
\beq
Q_e = - \kappa \left[N\right]^{x=+\infty}_{x=-\infty} \equiv - \kappa \delta N.
\label{eq:e}
\eeq
Thus any solitons with $\delta N \neq 0$ have necessarily non zero electric charge.

From Eq.~(\ref{eq:bb_bf}), we can put a BPS-like bound from blow on the energy $E = \int dx\, {\cal H}$
for a given charges $T_M$, $T_N$ $Q_e$ and $Q$ as
\beq
E \ge \sqrt{
\left(T_M + T_N + \delta n Q + \bar n Q_e \right)^2
+ \left(\delta m Q + \bar m Q_e\right)^2
},\quad
\tan\alpha = \frac{\delta m Q + \bar m Q_e}{
T_M + T_N + \delta n Q + \bar n Q_e},
\label{eq:bf_mass}
\eeq
with $\bar m \equiv (m_1+m_2)/2$ and $\delta m \equiv m_2 -m_1$ and the same for $n_i$.
Here we made use of
\beq
\left[
\sum_{i=1,2} m_i |\phi_i|^2
\right]^{+\infty}_{-\infty}
= \kappa \left[ M^2 \right]^{+\infty}_{-\infty},\quad
\left[
\sum_{i=1,2} n_i |\phi_i|^2
\right]^{+\infty}_{-\infty}
= \kappa \left[ MN \right]^{+\infty}_{-\infty}.
\eeq
Thus the energy and $Q$-charge are summarized as
\beq
E &=& \left(T_M + T_N + \delta n Q + \bar n Q_e \right) \frac{1}{\cos\alpha},\\
Q &=& \frac{(T_M+T_N)\tan\alpha - (\bar m - \bar n \tan\alpha)Q_e}{\delta m - \delta n \tan\alpha}.
\label{eq:q}
\eeq
We will call the kink with tricharges (topological, $Q$ and electric charges) trion in 1+1 dimensions.

The Bogomolnyi bound is saturated when the first two lines 
in Eq.~(\ref{eq:bb_bf}) vanish. For simplicity, let us 
introduce new variables 
\beq
\left(
\begin{array}{c}
\hat M\\
\hat N
\end{array}
\right) = 
U_\alpha \left(
\begin{array}{c}
M\\
N
\end{array}
\right),\quad
\left(
\begin{array}{c}
\hat m_i\\
\hat n_i
\end{array}
\right) = 
U_\alpha \left(
\begin{array}{c}
m_i\\
n_i
\end{array}
\right),\quad
U_\alpha \equiv 
\left(
\begin{array}{cc}
\cos\alpha & - \sin\alpha\\
\sin\alpha & \cos\alpha
\end{array}
\right).
\eeq
Then we are left with the self-dual equations
\beq
\D_1 \phi_i &=& - \left(\hat M - \hat m_i\right) \phi_i,
\label{eq:bf_bps1}\\
\D_0 \phi_i &=& i \left(\hat N - \hat n_i\right) \phi_i.
\label{eq:bf_bps2}
\eeq
Eq.~(\ref{eq:bf_bps2}) is solved by
\beq
A_0 = \hat N,\quad \phi_i(t,x) = e^{-i\hat n_i t} \tilde \phi_i(x).
\eeq
Let us rewrite the fields by
\beq
\tilde \phi_i = e^{-\frac{\psi(x)}{2}} \phi_{0i}(x) e^{\hat m_i x},\quad
\frac{\psi'}{2} = \hat M + i A_1.
\label{eq:bf_mm}
\eeq
Note that there is an equivalence relation 
\beq
\left(\phi_{0i},e^{\frac{\psi}{2}}\right) \sim \lambda
\left(\phi_{0i},e^{\frac{\psi}{2}}\right),\quad
\lambda \in \mathbb{C}^*.
\label{eq:V}
\eeq
This does not affect the physical fields in Eq.~(\ref{eq:bf_mm}).
Plugging these into Eq.~(\ref{eq:bf_bps1}), we get 
$\p_1 \phi_{0i} = 0$. 
Thus $\phi_{0i}$ is nothing but an integration constant, namely a zero mode of the solution.
We call it the moduli matrix by analogy with the domain wall solutions in ${\cal N}=2$
supersymmetric Yang-Mills theory \cite{Isozumi:2004jc,Isozumi:2004va}.
The last task is to determine $\psi$ for a given  set of $\NF=2$ constants $\{\phi_{0i}\}$.
The Gauss' law (\ref{eq:bf_gauss}) and the definition (\ref{eq:def_M}) lead to
\beq
\kappa N' &=& - \sum_{i=1,2} J_{0i} = 2 \sum_{i=1,2} \left( \hat N - \hat n_i\right) |\phi_i|^2,
\label{eq:n1}\\
\kappa M' &=& \sum_{i=1,2} (|\phi_i|^2)' = -2 \sum_{i=1,2} \left( \hat M - \hat m_i\right) |\phi_i|^2.
\label{eq:m1}
\eeq
This can be put together in the following single equation
\beq
\kappa \hat M' = -2 \sum_{i=1,2} \left(M-m_i\right)|\phi_i|^2.
\label{eq:m2}
\eeq
This is the equation for a gauge singlet object
\beq
\omega \equiv \frac{\psi + \psi^*}{2},\quad
\omega' = 2 \hat M.
\label{eq:omega}
\eeq
It follows that
\beq
\frac{\kappa^2}{4} \omega''
= \sum_{i=1,2} 
\left(
m_i\kappa - \sum_{j=1,2} e^{-\omega + 2\hat m_j x} |\phi_{0j}|^2
\right) e^{-\omega + 2\hat m_ix}|\phi_{0i}|^2.
\label{eq:bf_master}
\eeq
We call this the master equation again by analogy with 
the domain walls \cite{Isozumi:2004jc,Isozumi:2004va}.

Let us study the vacuum configurations with respect to the moduli matrix (\ref{eq:bf_mm})
with $\alpha = 0$. 
The $i$-th Higgs vacuum in Eq.~(\ref{eq:bf_higgs}) is given by
\beq
\text{Higgs}\ \left<i\right>\ :\ 
\omega = 2 m_i x + \log \frac{|a_i|^2}{m_i\kappa} ,\quad 
\phi_{0j} = a_i\delta_{ij}.
\label{eq:weight_h}
\eeq
The complex parameter $a_i$ can be set to $a_i=1$ by using the equivalence relation (\ref{eq:V}).
This solution $\omega$ for each vacuum is called the weight of vacuum~\cite{Eto:2005fm,Eto:2005cp}.
The Coulomb vacua (\ref{eq:bf_coulomb}) is given by
\beq
\text{Coulomb}\ \left<0\right>\ :\ \omega = 2 m_0 x + \log a_0^2,\quad {}^{\forall}\phi_{0i} = 0.
\label{eq:weight_c}
\eeq
Here $a_0,m_0$ are arbitrary real constants. 
Only $a_0$ can be fixed by the equivalence relation (\ref{eq:V}).

There are three different topological solitons which are kinks interpolating 
$\left<0\right>$-$\left<1\right>$,
$\left<0\right>$-$\left<2\right>$ and
$\left<1\right>$-$\left<2\right>$ vacua. Furthermore, there also exist
kinks connecting two points 
$\left<0\right>$-$\left<0\right>$ in the Coulomb vacua which are called non-topological solitons
\cite{Kao:1996tv}. We are interested in the kink $\left<1\right>$-$\left<2\right>$,
namely the trions, 
in this paper\footnote{
The kinks $\left<0\right>$-$\left<1\right>$ and 
$\left<0\right>$-$\left<0\right>$ were studied in \cite{Kao:1996tv}.}.
Thus we solve the master equation (\ref{eq:bf_master}) 
with the boundary condition at $x=\pm \infty$
\beq
\omega \to 2\hat m_i x + \log \frac{|a_i|^2}{m_i\kappa}.
\label{eq:bc}
\eeq
The moduli parameter of the trion solution $\left<1\right>$-$\left<2\right>$ is easily read
from the moduli matrix
\beq
(\phi_{01},\phi_{02}) = (a_1,a_2),\quad a_1,a_2 \in \mathbb{C}^*.
\label{eq:hw}
\eeq
One of $a_i$ can be fixed by the equivalence relation (\ref{eq:V}), so that the moduli space 
of the trion $\left<1\right>$-$\left<2\right>$ is determined as
\beq
{\cal M}^{\left<1\right>\text{-}\left<2\right>} = \mathbb{C}^* \simeq \mathbb{R} \times S^1.
\eeq
Here $\mathbb{R}$ is the position which is related to the broken translational symmetry while
$S^1$ is the Nambu-Goldstone mode associated with the broken global $U(1)$ symmetry.
The position can be estimated without solving Eq.~(\ref{eq:bf_master}) just by equating
two weights of vacua \cite{Eto:2006pg} as
\beq
x \simeq 
\frac{1}{\hat m_1 - \hat m_2} \left(\log \frac{|a_2|}{|a_1|} - \frac{1}{2}\log \frac{m_2}{m_1}\right).
\label{eq:estimate}
\eeq

\begin{figure}[t]
\begin{center}
\includegraphics[width=17.5cm]{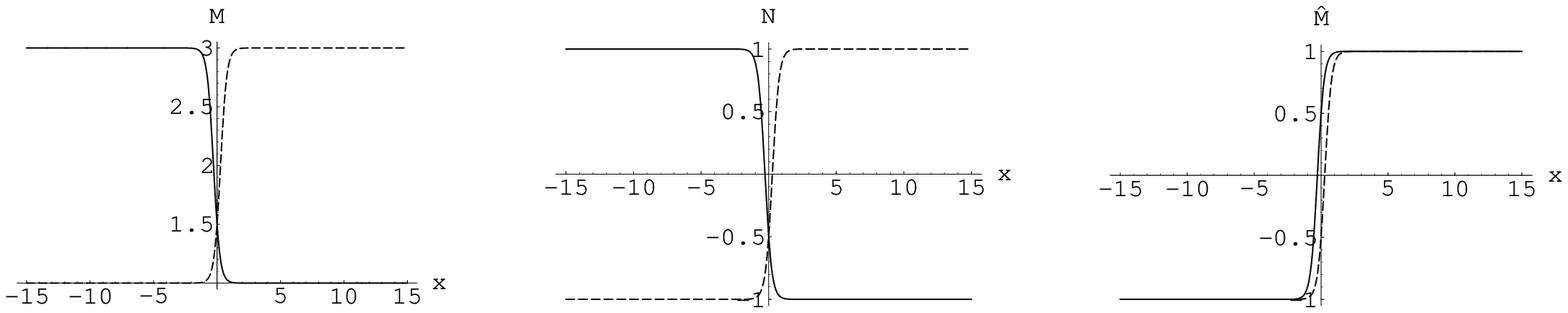}
\caption{{\footnotesize The numerical solutions  with $m=1$. The solid lines are
the trion ($\alpha = \pi/2$) and the dashed lines are the anti-trion ($\alpha=-\pi/2$).
}}
\label{fig:mn}
\end{center}
\end{figure}

To be more concrete, let us study a specific case with the following masses
\beq
(m_1,n_1) = \left(3m, m\right),\quad
(m_2,n_2) = \left(m, -m\right),
\label{eq:case2}
\eeq
with a non-negative constant $m$.
Then let us solve the master equation (\ref{eq:bf_master}) for a particular choice of the moduli matrix
\beq
(\phi_{01},\phi_{02}) = \left(\frac{1}{\sqrt{2}},\frac{1}{\sqrt{2}}\right),
\label{eq:mm_ex}
\eeq
and with the corresponding boundary condition (\ref{eq:bc}).
The trion position is estimated through Eq.~(\ref{eq:estimate}).
Note that $\hat m_1 - \hat m_2$ changes its sign as
\beq
\hat m_1 \ge \hat m_2\quad &\text{for}& \quad \alpha \in \left[-\frac{3\pi}{4},\frac{\pi}{4}\right],\\
\hat m_2 \ge \hat m_1\quad &\text{for}& \quad \alpha \in \left[-\pi,-\frac{3\pi}{4}\right],\ \left[\frac{\pi}{4},\pi\right].
\eeq
An orientation of trion depends on $\hat m_1 - \hat m_2$. We will call
the solutions for $\hat m_1 - \hat m_2 > 0$ trion while those for
$\hat m_1 - \hat m_2 < 0$ anti-trion.
In Fig.~\ref{fig:mn}, we show two solutions; one is the trion with $\alpha=\pi/2$ and the other 
is the anti-trion with $\alpha = -\pi/2$.
Note that the orientation of $\hat M$ is always same, see the right-most panel of
Fig.~\ref{fig:mn}.
Because the electric charge depends on $\delta N$ in Eq.~(\ref{eq:e}), the trion and anti-trion
have the electric charges with opposite sign.
The charges are explicitly given by
\beq
T_{M} = 4\kappa m^2,\quad T_N = 0,\quad
Q_e = \pm 2 m \kappa,
\eeq
where the upper sign is for trion and the lower for anti-trion. 
The $Q$-charge can be read  from Eq.~(\ref{eq:q}).

In the original Lagrangian (\ref{eq:bf}),
we can add an $\theta$-term
\beq
{\cal L}_{\theta} = \theta F_{01}.
\eeq
However, the $\theta$-term can be absorbed in the BF term by shifting $N$ and $n_i$ as
$N \to N+\theta/\kappa$ and $n_i \to n_i + \theta/\kappa$.
Since $T_N + \bar n Q_e$ is invariant under the above shift, the $\theta$-term
does not play any role classically for the trions.
In other words, the overall shift of $n_i$ is unphysical.

On the contrary a shift $m_i \to m_i + v^2/\kappa$ (a non-negative constant $v^2$)
is physically non-trivial. 
Then the energy is changed as
\beq
E = \sqrt{
\left(T_M + {\cal T}_M +T_N + \delta n Q + \bar n Q_e \right)^2
+ \left( -{\cal T}_N + \delta m Q + \bar m Q_e\right)^2
},
\label{eq:tension2}
\eeq
with another type of topological charges
\beq
{\cal T}_X = v^2 \int dx\ \p_1 X,\quad (X = M,N).
\eeq
Clearly, the above mass shift can be think of as the shift $M \to  M - v^2/\kappa$.
The definition of $M$ is slightly changed from Eq.~(\ref{eq:def_M}) as
\beq
\kappa M \equiv |\phi_1|^2 + |\phi_2|^2 - v^2.
\label{eq:mshift}
\eeq
This redefinition is useful when we compare the trions and the $Q$-kinks~\cite{Abraham:1992vb,Abraham:1992qv}.
To this end, let us take $\kappa \to 0$ limit with $v^2$ and $m_i$ being fixed ($v_i^2 = \kappa m_i \to 0$). 
In the limit, Eq.~(\ref{eq:mshift}) becomes a constraint on $\phi_1,\phi_2$ which
forces $\phi_1,\phi_2$ take the value in $S^3$. Taking the $U(1)$ gauge symmetry into
account, the BF theory with $\kappa=0$ is a massive non-linear sigma model whose
target space is $\mathbb{C}P^1 \simeq S^3/S^1$. The parameter $v^2$ is related to the
radius of $\mathbb{C}P^1$. Thus the trions in the BF theory
continuously go to the  $Q$-kinks in the $\mathbb{C}P^1$ model.
For the $Q$-kinks, the master equation (\ref{eq:bf_master})
is no longer differential equation, so that
we can analytically solve it~\cite{Isozumi:2004jc,Isozumi:2004va} as
\beq
\omega = \log \left( v^{-2} \sum_{i=1,2}e^{2\hat m_i x}|\phi_{0i}|^2\right),
\eeq
and $M,N$ are also analytically obtained as
\beq
M = \frac{\sum_{i=1,2}m_i e^{-\omega +2\hat m_ix}|\phi_{0i}|^2}{\sum_{i=1,2} e^{-\omega +2\hat m_ix}|\phi_{0i}|^2},\quad
N = \frac{\sum_{i=1,2}n_i e^{-\omega +2\hat m_ix}|\phi_{0i}|^2}{\sum_{i=1,2} e^{-\omega +2\hat m_ix}|\phi_{0i}|^2}.
\eeq

\begin{figure}[t]
\begin{center}
\includegraphics[width=8cm]{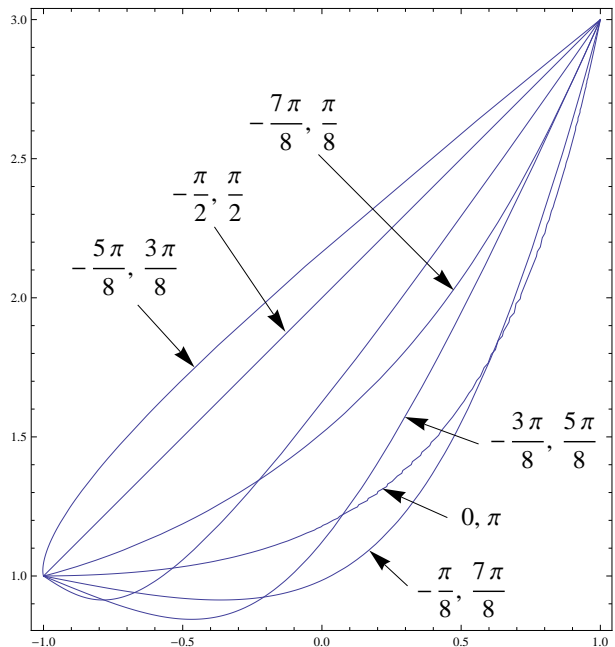}
\caption{{\footnotesize The trajectories on the $NM$ plane. 
Trions ($\alpha = 3\pi/8,\pi/2,5\pi/8,3\pi/4,7\pi/8,\pi$) run from $(m,3m)$
to $(-m,m)$ while anti-trions ($\alpha = - 5\pi/8,-\pi/2,-3\pi/8,-\pi/4,0$) run 
from $(-m,m)$ to $(m,3m)$ with $m=1$.
}}
\label{fig:trajectry}
\end{center}
\end{figure}
We are ready to compare the $Q$-kinks and the trions. 
First, the $Q$-kinks are electrically neutral while the trions are electrically
charged. 
Its mass is obtained from Eq.~(\ref{eq:tension2})
\beq
E_{Q} = \sqrt{
\left({\cal T}_M + \delta n Q \right)^2
+ \left( -{\cal T}_N + \delta m Q\right)^2
}.
\eeq
${\cal T}_{M,N}$ is the topological charge of $\mathbb{C}P^N$ kinks.
Second, trajectories of $Q$-kinks on the $NM$ plane are always straight segments
between $(n_1,m_1)$ and $(n_2,m_2)$.
On the other hand, the trajectories of trions are not always straight.
Generally, they are curved segments connecting $(n_1,m_1)$ and $(n_2,m_2)$.
We show several trajectories of trions for the masses choice (\ref{eq:case2}) in 
Fig.~\ref{fig:trajectry}. The trions only for $\alpha = \pm \pi/2$ are straight segments while
all the others are curved trajectories.
Thus the trions which we have found in this letter are quite different from the $Q$-kinks.\\

\noindent{\bf Concluding Remarks}

We find new topological solitons in the $1+1$ dimensional BF Higgs theory.
They have the topological charge, $Q$-charge and electric charge. Hence we call them the trions.
We have derived the BPS-like energy bound and solved the self-dual equations. We also have
found all the zero modes of the single trion. We have found that the trions are quite different
from the ordinary $Q$-kinks which are electrically neutral.

There are several directions to study the trions.
It was pointed out that the dyons and the $Q$-kinks in two dimensions
share many similar properties \cite{Abraham:1992vb,Abraham:1992qv}.
If we pursue the similarity for trions, we may expect existence of new topological solitons with
tri-charges in four dimensions.
One possible scenario is to seek them in the Higgs phase. Recently, a confined monopole in the Higgs
phase was identified with a kink inside the non-Abelian vortex in ${\cal N}=2$ supersymmetric
gauge theory in four dimensions \cite{Tong:2003pz}. 
If the BF coupling is induced by radiative 
quantum effects on the effective vortex world-sheet theory, we would be able to construct the trions inside the vortex. 
From the four dimensional view point, it is the topological soliton with the tri-charge confined inside the flux tube.
It is also interesting to find multiple trions. In our model 
there exists only one trion since we have only two Higgs vacua. In models with more scalar
fields $\NF > 2$, we have $\NF$ Higgs vacua and there may exist $\NF-1$ multiple trions in
such models. Scattering of multiple trions is also an interesting problem.
Moreover, it may be interesting to explore trions in non-relativistic BF theories
and in non-Abelian BF theories.\\\


The author thank to Koji Hashimoto, Takaaki Ishii, Muneto Nitta and Ta-sheng Tai for
useful comments.
This work is supported by Special Postdoctoral Researchers Program at RIKEN.

\end{document}